\documentclass[12pt,preprint]{aastex}

\slugcomment{Submitted to Astrophysical Journal Letters}

\shorttitle{RESIK observations in solar flares} \shortauthors{Sylwester et al.}

\begin{document}


\title{RESIK OBSERVATIONS OF HELIUM-LIKE ARGON X-RAY LINE EMISSION IN SOLAR FLARES}

\author{J. Sylwester and B. Sylwester}
\affil{Space Research Centre, Polish Academy of Sciences, 51-622, Kopernika~11, Wroc{\l}aw, Poland}
\email{js@cbk.pan.wroc.pl}

\and

\author{K. J. H. Phillips}
\affil{UCL-Mullard Space Science Laboratory, Holmbury St Mary, Dorking, Surrey, RH5 6NT, United
Kingdom} \email{kjhp@mssl.ucl.ac.uk}

\begin{abstract}
The \ion{Ar}{17} X-ray line group principally due to transitions $1s^2 - 1s2l$ ($l=s$, $p$) near
4~\AA\ was observed in numerous flares by the RESIK bent crystal spectrometer aboard {\it
CORONAS-F} between 2001 and 2003. The three line features include the \ion{Ar}{17}  $w$ (resonance
line), a blend of $x$ and $y$ (intercombination lines), and $z$ (forbidden line), all of which are
blended with \ion{Ar}{16} dielectronic satellites. The ratio $G$, equal to $[I(x)+I(y) +
I(z)]/I(w)$, varies with electron temperature $T_e$ mostly because of unresolved dielectronic
satellites. With temperatures estimated from {\it GOES} X-ray emission, the observed $G$ ratios
agree fairly well with those calculated from {\sc chianti} and other data. With a two-component
emission measure, better agreement is achieved. Some \ion{S}{15} and \ion{S}{16} lines blend with
the Ar lines, the effect of which occurs at temperatures $\gtrsim 8$~MK, allowing the S/Ar
abundance ratio to be determined. This is found to agree with coronal values. A nonthermal
contribution is indicated for some spectra in the repeating-pulse flare of 2003 February~6.

\end{abstract}


\section{INTRODUCTION}

High-resolution soft X-ray spectral observations have provided important information about the
physical properties of solar flares and active regions, including electron temperatures $T_e$,
densities $N_e$, and ionization state. Those from the bent crystal spectrometer RESIK (REntgenovsky
Spektrometr s Izognutymi Kristalami) on the Russian {\it CORONAS-F} spacecraft \citep{syl05a} cover
the range 3.3--6.1~\AA, which includes strong emission lines of Si, S, and Ar ions and continuum,
and weaker lines of Cl and K ions. The wavelength resolution $\Delta \lambda$ is between 8~m\AA\
(3.4~\AA) and 17~m\AA\ (6.1~\AA). The instrument operated successfully between 2001 (shortly after
the launch of {\it CORONAS-F} on July~31) and May 2003, obtaining numerous flare and active region
spectra. RESIK spectra have already been used for K, Ar, Cl, S, and Si abundance determinations in
flares \citep{syl06}. Lines in the 3.3--6.1~\AA\ range have not been well observed by previous
spectrometers with the exception of the He-like S (\ion{S}{15}) lines at $\sim 5$~\AA. Of
particular interest here are the He-like Ar (\ion{Ar}{17}) lines between 3.94~\AA\ and 4.00~\AA\
and associated dielectronic satellites emitted by Li-like Ar (\ion{Ar}{16}). There are four
prominent \ion{Ar}{17} lines, due to transitions from the $1s2p\,^1P_1$, $1s2p\,^3P_2$,
$1s2p\,^3P_1$, and $1s2s\,^3S_1$ levels to the ground level $1s^2\,^1S_0$ (designated lines $w$,
$x$, $y$, and $z$ respectively), with wavelengths 3.949~\AA, 3.966~\AA, 3.969~\AA, and 3.994~\AA.
Dielectronic \ion{Ar}{16} satellites, particularly $q$ (3.981~\AA), $k$ (3.990~\AA), and $j$
(3.994~\AA) feature prominently.

The \ion{Ar}{17} lines have been observed during a few solar flares with the Flat Crystal
Spectrometer (FCS) on {\it Solar Maximum Mission}, with the \ion{Ar}{16} $k$ satellite being
resolved from \ion{Ar}{17} line $z$ but $j$ indistinguishable from line $z$. RESIK flare spectra do
not resolve lines $k$ and $z$ or the \ion{Ar}{17} $x$ and $y$ lines, so there are three main
features, the blend of line $w$ with unresolved weak \ion{Ar}{16} satellites on the long-wavelength
side of $w$ ($w'$), the blend of lines $x$, $y$, and other weak \ion{Ar}{16} satellites [$(x+y)'$],
and the blend of line $z$ with the two prominent satellites $j$ and $k$ ($z'$). The relative fluxes
of these three line features can thus be examined over a variety of flare conditions, RESIK's
sensitivity being much improved (by a factor $\sim 60$) over the FCS. In particular, the
temperature dependence of the ratio $G$, which in terms of the observed features is $[I(x+y)' +
I(z')]/I(w')$, can be examined. At temperatures $\gtrsim 8$~MK, the effect of sulphur lines
blending with the three \ion{Ar}{17} line features becomes apparent on the $G$ ratio. These are the
$1s-3p$ (Ly-$\beta$) line of \ion{S}{16} (3.991~\AA) and $1s^2-1s5p$ line of \ion{S}{15}
(3.998~\AA), blending with the \ion{Ar}{17} $z$ line, and the $1s^2-1s6p$ line of \ion{S}{15}
(3.949~\AA) blending with the \ion{Ar}{17} $w$ line. These lines were not considered in an earlier
analysis of FCS spectra \citep{phi93}.

\section{OBSERVATIONS}

\begin{figure}
\epsscale{1.5}
\begin{center}
\plottwo{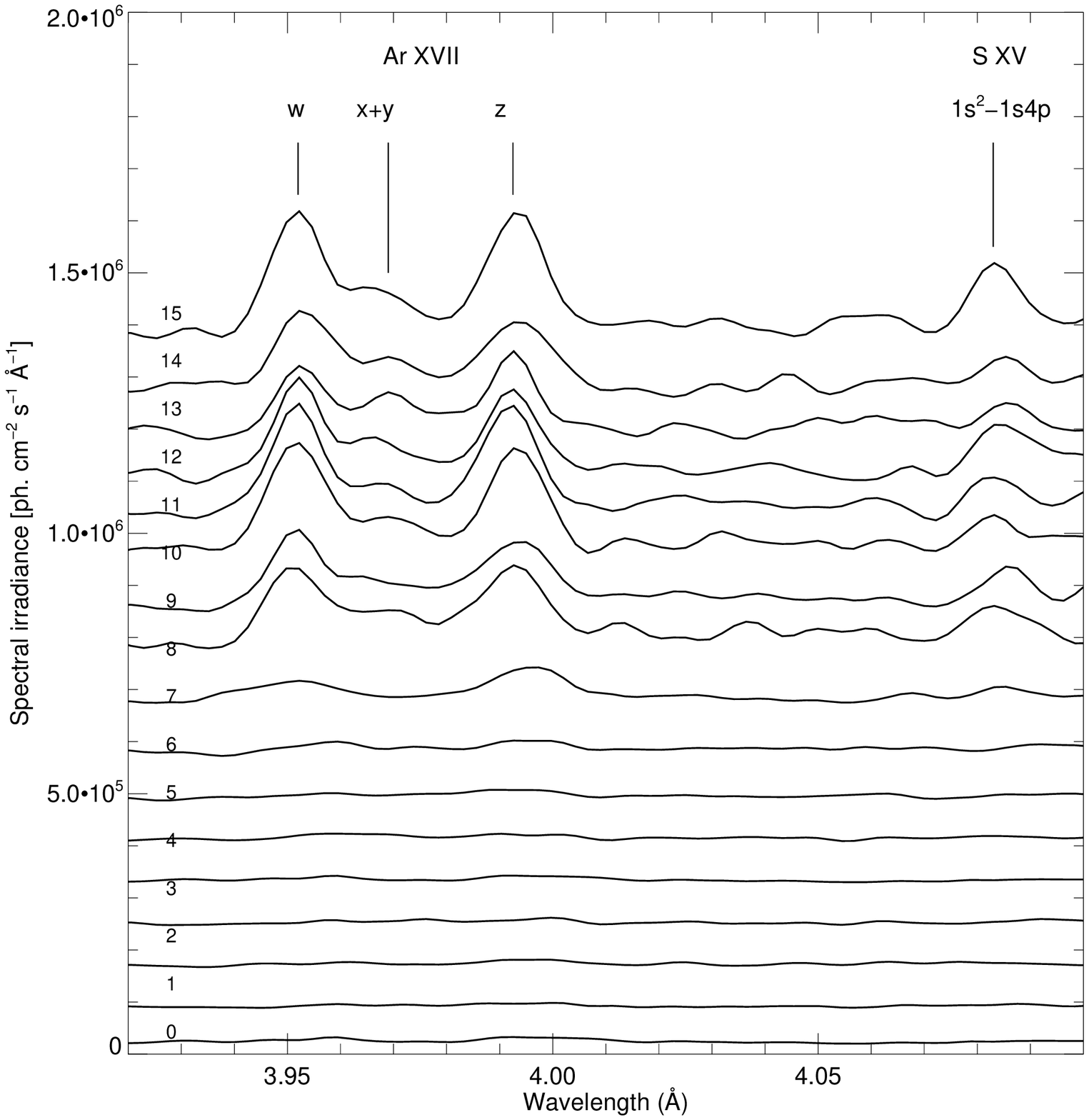}{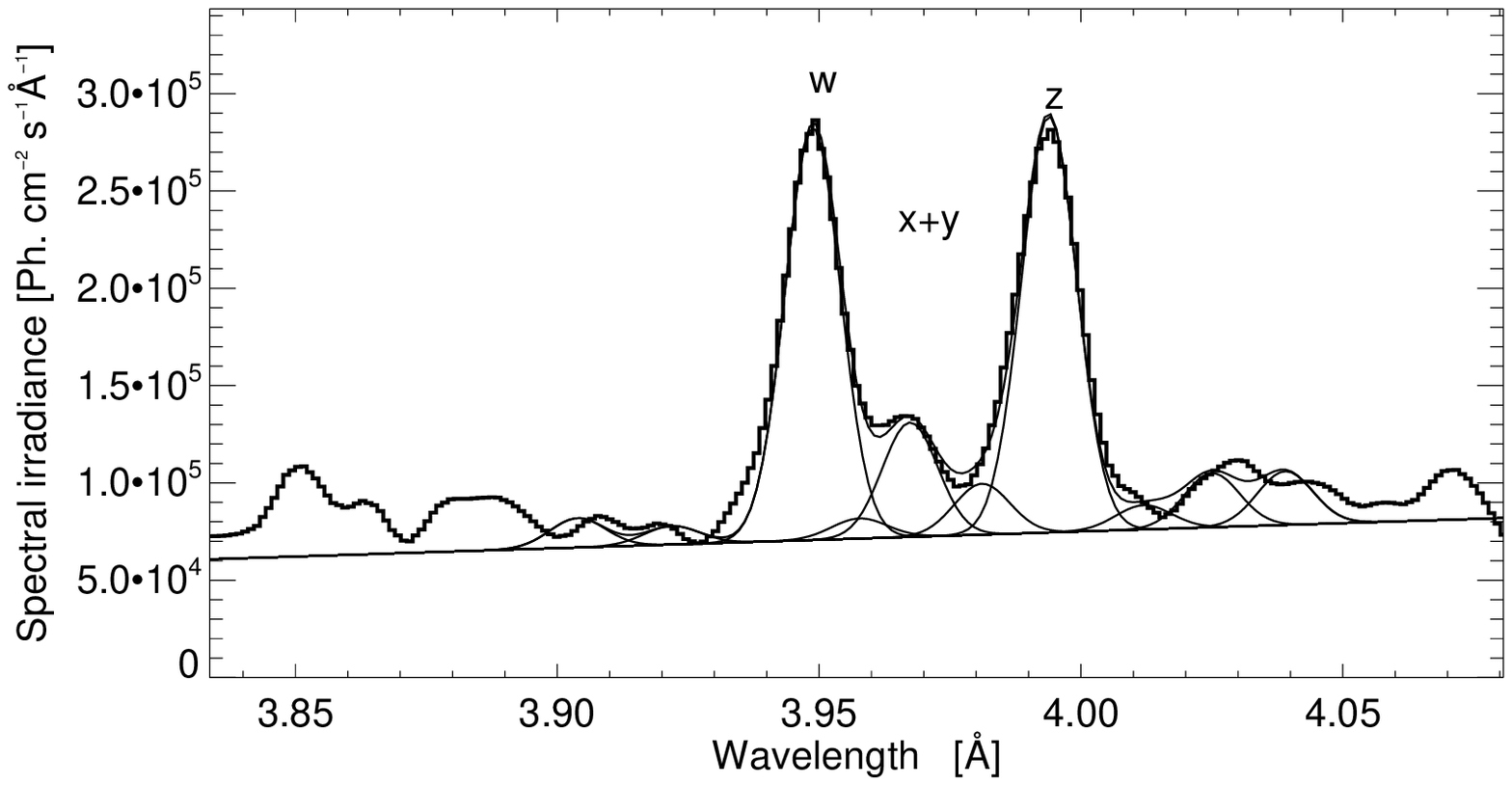} \caption{{\it Upper:} Examples of RESIK spectra in the 3.85--4.07~\AA\
region including the \ion{Ar}{17} lines described in the text and the \ion{S}{15} $1s^2 - 1s4p$
line at 4.089~\AA. They cover the period 0200:44--0212:08~UT extending over the rise phase to the
first maximum of the 2003 February~6 flare. Spectra are stacked with times increasing upwards. {\it
Lower:} Gaussian fits to the RESIK spectrum at 02:10:47--02:11:03~UT (no. 11) illustrating the line
fitting procedure (with the IDL routine lmfit). Only the line features $w$, $x+y$, and $z$ are
real. } \label{RESIK_spectra}
\end{center}
\end{figure}

The observations analyzed here were made with channel~2 of RESIK (nominal range for an on-axis
source 3.83--4.27~\AA), for which the diffracting crystal is Si 111 ($2d = 6.27$~\AA). A 0.5\%
background due to fluorescence of the crystal material for this channel has been completely
accounted for. During a flare, spectra are recorded in data gathering intervals (DGIs) that are
adjustable according to the X-ray flux. For a typical {\it GOES} class M1 flare, the DGI may be
from several minutes at the start and end of the flare to $\sim 2$~s at flare peak. Some 1500
spectra during seven flares of various importance are available, from which 98 flare maxima or
post-maxima spectra were selected and analyzed in detail.  Details of the time periods and flare
importance are given in Table~\ref{Spectra_list}. These spectra have been processed to account for
all known instrumental effects.\footnote{The data are made available to the solar physics community
through the World Wide Web site http://www.cbk.pan.wroc.pl/RESIK\_Level2 based at the Space
Research Centre, Wroc{\l}aw, Poland.} Spectra during the rise of the flare of 2003 February~6 are
shown in Figure~\ref{RESIK_spectra} (upper panel). Gaussian profiles to the main line features
$w'$, $(x+y)'$, and $z'$ were fitted using a nonlinear least-squares fitting procedure with an
arbitrary number of parameters available in a library of Interactive Data Language (IDL)
procedures. The output consists of estimated fluxes with uncertainties of the three observed
features from which $G$ ratios with uncertainties can be derived. Fits to one of the spectra in the
upper panel of Figure~\ref{RESIK_spectra} are shown in the lower panel.

Temperatures over the period of each spectrum were estimated from the flux ratio of the two
channels (0.5--4~\AA\ and 1--8~\AA) of {\it GOES}. In this procedure, standard IDL routines were
used to derive the temperatures which are based on the flux sensitivity of the {\it GOES} channels
and model X-ray spectra calculated with version~5.2 of the {\sc chianti} atomic database and
spectral code \citep{lan06}. The routines are from \cite{whi05}, in which account is taken of
differing coronal and photospheric element abundances. Temperatures were derived here using coronal
abundances (later we show that the coronal Ar/S abundance ratio fits our data better) but for the
temperature range of RESIK spectra analyzed in this work (5--16~MK) the temperature differences
using coronal and photospheric abundances are very small, between 0.9~MK and 0.5~MK, in agreement
with \cite{whi05}.

As well as RESIK spectra, four {\it SMM}\/ FCS spectra during flares in 1988 were included in the
analysis, as listed by \cite{phi93}. As the spectra were scanned in each case, the different
\ion{Ar}{17} lines were observed at slightly different times, unlike the RESIK spectra which are
formed simultaneously, but the scan times were only 3 minutes, short enough that insignificant time
variations occurred. In addition, a laboratory spectrum of the \ion{Ar}{17} lines \citep{bit03} was
included in the data. The spectrum in this case was taken during Ohmically heated discharges from
the National Spherical Torus Experiment (NSTE), with temperature measured from Thomson scattering.
The electron densities are much higher ($5\times 10^{13}$~cm$^{-3}$) than those expected in flares,
but the comparison of the $G$ ratio should still be valid as it is a function of $T_e$ alone.

\section{ATOMIC CALCULATIONS}

For He-like ion lines, the ratio $G = [I(x) + I(y) + I(z)]/I(w)$ is slightly sensitive to electron
temperature $T_e$ owing to the different energy dependence of the collisional excitation rate
coefficients from the ground state to the upper levels of each line \citep{gab69}.  The theoretical
temperature dependence of the \ion{Ar}{17} $G$ ratio used here is based on data in the {\sc
chianti} atomic database (v.~5.2), taken from \cite{zha87}. This is shown in Figure~\ref{G-ratio}
({\it lower panel}, solid line). The effect of \ion{Ar}{16} dielectronic satellites on the $G$
ratio is of major consequence. Satellites with transitions $1s^2\,nl - 1s2l\,nl'$ ($n\geq 3$)
converge on and blend with lines $w$ and $y$, while the prominent satellites $j$ and $k$ (with
$n=2$) blend with line $z$. The contribution functions (fluxes per unit emission measure) of all
the $n=2$ and $n= 3$ satellites for different temperatures were taken from {\sc chianti}. The
combined effect of higher-$n$ satellites ($n\geqslant 4$) is also important, but in the absence of
any available data, their contribution was estimated from factors for equivalent \ion{Fe}{24}
satellites \citep{bel79}. The effect of all \ion{Ar}{16} satellites on the observed $G$ ratio
($G'$) is shown in Figure~\ref{G-ratio} ({\it lower panel}, dashed line). Note that for much of the
temperature range shown, $\log T_e = 6.7-7.2$, the He-like stage Ar$^{+16}$ dominates over all
other Ar ionization stages \citep{maz98}.

The effect on the $G$ ratio of \ion{S}{15} and \ion{S}{16} lines was estimated from {\sc chianti}.
The \ion{S}{16} $1s-3p$ (Ly-$\beta$) line at temperatures $\gtrsim 8$~MK, blending with the
\ion{Ar}{17} $z$ line, is particularly important, tending to increase $G$. The exact effect of the
S lines on $G$ depends on an assumed S/Ar abundance ratio, which for coronal abundances
\citep{fel92} is 4.9 but for photospheric abundances \citep{gre98} is 8.5. The dash--treble-dot
($G''_{\rm cor}$) and dash--dot curves ($G''_{\rm phot}$) show the effect of the S lines on the $G$
for coronal and photospheric abundances respectively. The difference between the two curves is
sufficiently large at $T_e \gtrsim 12$~MK that RESIK flare spectra can be expected to distinguish
between coronal and photospheric abundances for flares.

\section{RESULTS}

\begin{figure}
\epsscale{.6}
\begin{center}
\plotone{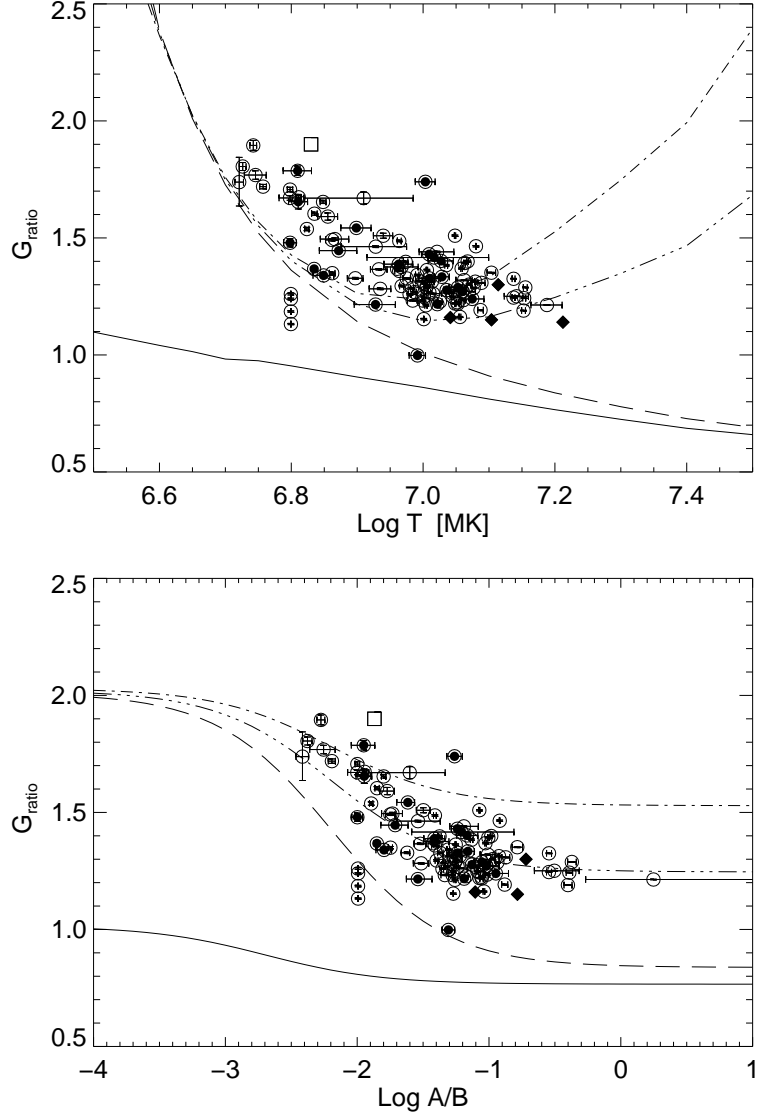} \caption{{\it Upper:} The \ion{Ar}{17} $G$ ratio plotted against $\log T_e$ in an
isothermal approximation. For theoretical calculations, the solid line is $G = [I(x) + I(y) +
I(z)]/I(w)$ for the \ion{Ar}{17} lines alone; dashed line is $G$ including \ion{Ar}{16} satellites;
dash--treble-dot line is $G$ including \ion{S}{15} and \ion{S}{16} lines for coronal Ar/S abundance
($G''_{\rm cor}$), and dash--dot line includes the S lines with photospheric Ar/S abundance
($G''_{\rm phot}$). Observational points (circles with small error bars) are measured $G$ for
spectra during flares (Table~\ref{Spectra_list}), with filled circles for spectra during the 2003
February~6 flare. The filled diamond symbols are for the {\it SMM}\/ FCS spectra. Temperatures $T$
are electron temperatures for the theoretical curves and for the observational points are derived
from the ratio of the two {\it GOES} channels. The square symbol is for the NSTE tokamak
\ion{Ar}{17} spectrum, with temperatures from Thomson scattering \citep{bit03}. {\it Lower:} The
\ion{Ar}{17} $G$ ratio plotted against $\log (B/A)$ for a two-component DEM $= A\delta(T_1) +
B\delta(T_2)$ with $T_1=4.5$~MK and $T_2=16$~MK. The line styles and symbols for the observed
points have the same meaning. } \label{G-ratio}
\end{center}
\end{figure}

The observed $G$ ratios, defined by $G = [I(x+y)' + I(z')]/I(w')$ from estimated fluxes of the
\ion{Ar}{17} line features $w'$, $(x+y)'$, and $z'$, are plotted in Figure~\ref{G-ratio} ({\it
upper panel}) against temperatures from the flux ratio of the two {\it GOES} channels. The points
(including the RESIK non-flaring point, {\it SMM}\/ FCS points, and the NSTE tokamak point of
\cite{bit03}) follow a clear trend with temperature, though there is a general displacement of the
observed $G$s from the theoretical curves $G''_{\rm cor}$ (\ion{Ar}{17} $G$ ratio with S lines,
coronal Ar/S abundance ratio) such that either the {\it GOES} temperatures are too low (by 0.1 in
log~$T_{\rm GOES}$) or the ratios are $\sim 20$\% too high. A similar effect was obtained for
ratios of dielectronic satellites to \ion{Si}{13} $1s^2\,^1S_0 - 1snp\,^1P_1$ ($n=3$, 4) lines
observed by RESIK \citep{phi06}, which was attributed to non-isothermal plasma in the flare. In
general, a flare plasma is expected to have a differential emission measure (DEM) $\varphi(T_e) =
N_e^2 dV/dT_e$ extending over a broad temperature range, up to $T_e \sim 20\times 10^6$~K for M and
X flares. The value of $G$ can be derived from the dependence on $T_e$ of the DEM which in
principle can be found from several X-ray emission lines (e.g. \cite{syl80}, \cite{kep06}). In
previous work we have used simple forms for the DEM with some success, e.g. DEM = constant $\times
\exp (-T_e/T_0)$ ($T_0 = $ constant for each spectrum, calculable from the ratio of the emission in
the two {\it GOES} channels). This gave a fair agreement of observed and theoretical
\ion{Si}{12}/\ion{Si}{13} line ratios \citep{phi06}. For this analysis, however, analytic forms
were not so satisfactory, and instead, a two-component emission measure was found to give the best
agreement with the observations, with DEM given by the weighted sum of two delta functions, DEM $=
A\delta(T_1) + B\delta(T_2)$, the temperatures $T_1$ and $T_2$ being constant for all spectra but
$A$ and $B$ variable. This form of the DEM has been found to give satisfactory results for other
data \citep{syl05b}, and although its use here is more for convenience in the analysis, physical
importance might be attributed to the two temperatures in that $T_1$ appears to represent the
nonflaring active region temperature and $T_2$ the temperature of the flare proper. The theoretical
$G$ ratios, with and without the \ion{Ar}{16} satellites and the sulphur lines, are plotted against
$\log (B/A)$ in Figure~\ref{G-ratio} ({\it lower panel}) for the case of $T_1 = 4.5$~MK and $T_2 =
16$~MK. These values give the best agreement of the observed points and the theoretical curves; in
particular, the point of inflection between the low- and high-temperature asymptotic values of the
$G''_{\rm cor}$ curve best matches that in the observed points.

Values of $\log (B/A)$ for the observed $G$ ratios were obtained from the emission ratio of the two
{\it GOES} channels using the DEM delta-function form with $T_1 = 4.5$~MK and $T_2 = 16$~MK. These
are shown in Figure~\ref{G-ratio} ({\it lower panel}). There is now improved agreement with the
$G''_{\rm cor}$ curve (coronal S/Ar abundance ratio).  Values of $B/A$ range from 0.003 to 1.6,
i.e. DEMs with the $T_1 = 4.5$~MK component dominant to DEMs in which the two components are
comparable to each other.

For the 2003 February~6 flare, the \ion{Ar}{17} $x+y$ or $z$ line feature has anomalously high flux
compared with the $w$ line for two spectra, giving $G\sim 1.7$. For the $G''_{\rm cor}$ curve, this
is only expected for very low (5~MK) or high (18~MK) temperatures, but the {\it GOES} temperature
for these time periods are 8--10~MK. In Figure~\ref{G-ratio}, $G$ for these spectra are thus well
above the cluster of points. This flare is of particular interest because its light curve in X-rays
up to energies of 25~keV, as shown by {\it RHESSI} observations, consists of seven semi-periodic
pulses which originate in active regions either side of the solar equator, linked by a
transequatorial loop. \cite{fou05} discuss the excitation of individual pulses in terms of a fast
magnetoacoustic kink mode excited in the transequatorial loop. RESIK observations cover the period
02:03--02:17~UT, the period of the first two pulses. One of the anomalous $G$ ratios occurs at
02:09~UT when an impulsive burst in the {\it RHESSI} 12--25~keV range occurs, about 85~s preceding
the peak of the first pulse. The $G$ ratio is high because the $z'$ line feature is high relative
to the other \ion{Ar}{17} lines, and it is possible that $z'$ is enhanced because of the high
fluxes of the \ion{Ar}{16} $j$ and $k$ satellites which blend with the \ion{Ar}{17} $z$ line. The
upper levels of these lines are excited by electrons with energy 2.2~keV, so their enhancement may
be due to an overabundance of electrons with this energy. The time coincidence with the 12--25~keV
X-ray pulse would then suggest a nonthermal component of the \ion{Ar}{17} emission.

Both panels of Figure~\ref{G-ratio} indicate a much clearer agreement of the observed $G$ ratios
with the $G''_{\rm cor}$ curve, i.e. with coronal S/Ar abundance (4.9: \cite{fel92}), than the
$G''_{\rm phot}$ curve with photospheric S/Ar abundance (8.5: \cite{gre98}). This points to the
origin of the flare plasmas in the RESIK sample being coronal. According to \cite{fel00}, flare
plasmas have coronal abundances unless very impulsive when photospheric abundances appear to be
present. This accords with our analysis since the flares observed (Table~\ref{Spectra_list}) are
all gradual in character.

\section{CONCLUSIONS}

We have discussed RESIK observations of the $G$ ratio for the \ion{Ar}{17} lines at $\sim 4$~\AA\
over a wide range of activity. For 98 flare spectra as well as four {\it SMM}\/ FCS spectra, there
is a clear trend of $G$ that is close to the theoretical ratio. The agreement is improved for a
two-component DEM. From this it is deduced that the Ar/S abundance ratio is close to 4.9, its
coronal value, so indicating a coronal origin for the flare plasma. Two spectra with anomalous $G$
during the 2003 February~6 flare are noted, one of which is exactly at the time of an impulsive,
nonthermal burst in {\it RHESSI} 12--25~keV emission. There is thus some suggestion that part of
the \ion{Ar}{17} emission is nonthermal in origin.

\acknowledgments We thank the Polish Ministry of Science and Education for financial support from
Grant no. 1~P03D~017~29 and travel support from a U.K. Royal Society/Polish Academy of Sciences
International Joint Project. {\sc chianti} is a collaborative project involving NRL (USA), MSSL
(UK), the Universities of Florence (Italy) and Cambridge (UK), and George Mason University (USA).

\clearpage

\begin{deluxetable}{lccc}
\tabletypesize{\scriptsize} \tablecaption{A{\sc nalyzed} R{\sc esik} F{\sc lare} S{\sc pectra}
\label{Spectra_list}} \tablewidth{0pt}

\tablehead{\colhead{Date} & \colhead{Time Range of } & \colhead{Time of Maximum } & \colhead{{\it
GOES} Class}
\\
&\colhead{RESIK Spectra (UT)} & \colhead{{\it GOES} (1--8 \AA) Flux}  }

\startdata

2002 October 4 & 05:39--05:54 &  05:38 & M4.0 \\

2002 December 17 & 23:08--23:19 & 23:09 & C6.1 \\

2003 January 7 & 23:36--23:43 & 23:33 & M4.9 \\

2003 January 9  & 01:36--02:08  & 01:39 & C9.8 \\

2003 February 1 & 09:12--10:14 & 09:06 & M1.2 \\

2003 February 6 & 02:09--02:31 & 02:12 & C3.2 \\

2003 February 22 & 04:50--05:02 & 05:10 & B9.6 \\

\\

\enddata


\end{deluxetable}

\clearpage

\end{document}